\documentclass[a4paper,onecolumn]{agn6}
\usepackage{natbib}
\usepackage{graphicx}
\usepackage[a4paper]{hyperref}
\idline{}{1}
\begin{document}
   \title{Chandra observations of X-ray source overdensities in distant galaxy cluster fields
}
 \author{N.~Cappelluti \inst{1,5}
      \and
      M.~Cappi\inst{1}
      \and
      M.~ Dadina\inst{1}
      \and
      G.~Malaguti\inst{1}
     \and
      M.~Branchesi\inst{2,3}
     \and
      V. D'Elia \inst{4}
      G.G.C.~Palumbo\inst{2}}   
  \institute {{IASF-CNR, Sezione di Bologna, Via Gobetti 101, I-40129
               Bologna, Italy
               \email{cappelluti@bo.iasf.cnr.it} 
               \and
   Dipartimento di Astronomia Universit\`a di Bologna,
    via Ranzani 1, I--40127 Bologna, Italy
        \and 
      IRA-CNR, Via Gobetti 101, I-40129
               Bologna, Italy
		\and	
	INAF-Osservatorio astronomico di Roma, Via di Frascati, 33 I-00040 Monteporzio  Catone (Rome), Italy 
	\and
	 Max-Planck-Institute fuer Extraterrestrische Physik,  Giessenbachstrasse, 1  D--85741 Garching b. Munchen,  Germany                      
	}}

   \abstract{The first systematic study  of the serendipitous  X-ray source density around 
10 high-$z$ (0.24$<z<$1.2) clusters  performed with \emph{Chandra} is presented.
A factor $\sim2$ overdensity has been found in 4 cluster  fields.
The result is  statistically highly significant (at $>$2$\sigma$ per field, 
and $>$5$\sigma$ overall) only on scales 8$\arcmin$$\times$8$\arcmin$ 
and  peculiar to cluster fields. The blank field fluctuations, i.e. 
the most sensitive measurements to date of the X-ray cosmic variance on this 
scale, indicate (1$\sigma$) variations within $\sim$15-25\% around the average value.
The first marginal evidence that these overdensities increase 
with cluster redshift is also presented. 
We speculate that the most likely explanation is that 
the sources in excess are AGNs tracing filaments connected to the clusters. 
If the association of the overdensities with large-scale structures and their 
positive relation with cluster redshift is confirmed, these studies 
could represent  a new, direct, way of mapping the cosmic web and its cosmological 
parameters at high-$z$.
}
   \authorrunning{N. Cappelluti et al.}
   \titlerunning{X-ray source overdensities around distant clusters}
   \maketitle
%

\section{Introduction}

Recent studies have reported an unexpected excess of point-like X-ray sources in the outskirts of several distant clusters of galaxies 
(Henry \& Briel 1991, Cappi et al. 2001, Pentericci et al. 2002, Molnar et al. 2002, Johnson et al. 2003). 
The source number densities measured around the clusters have been found to exceed the value observed in 
similarly deep observations of fields without clusters by a factor $\sim$2.  This is in contrast with  optical and X-ray studies which  suggest that AGNs are somewhat rare in clusters environments both in the nearby Universe and at high-$z$.  
At least in one  case (3C295 at $z\sim$0.5), the excess number of sources has been demonstrated to be 
 asymmetric  with respect to the cluster, possibly indicating a filament of the large scale structure 
connected with the cluster (D'Elia et al. 2004). 
These pioneering studies  well emphasize the potential of  X-ray surveys for tracing the cosmic web at high-$z$. 
Within this framework we started a systematic and uniform 
 analysis of 
the X-ray source population around 10 distant clusters of galaxies.
 The study was performed using archival {\it Chandra} ACIS-I  observations which are well suited to detect
faint X-ray sources. Detailed study is shown in Cappelluti et al. (2004), the main results of which are summarized here.

\section {Sample selection and data analysis}
We selected from the Chandra public archive 10 observations of clusters of galaxies with $z>$0.2 in order to observe both the cluster  diffuse emissions and their outskirts. 
We selected as deep as possible  ACIS-I observations.
This choice permits  to analyze the X-ray source surface density with good statistics  over a wide field of view  (16$\arcmin \times 16\arcmin$).
We selected also 5 reference fields not centered on clusters with similar exposures  in order to compare the logN-logS over the 
same flux intervals.
  \begin{figure}
   \centering
    \includegraphics[width=6cm,angle=-90]{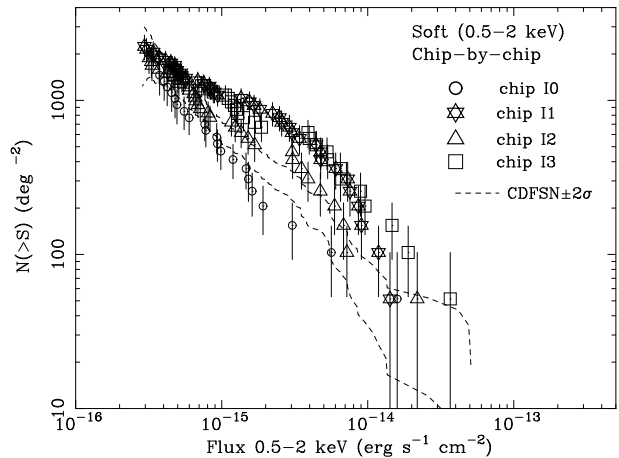}
      \caption{0.5-2 keV logN-logS calculated chip-by-chip in the field of MS 1137+6625 $z$=0.78             }
         \label{FigVibStab}
   \end{figure}
Data were reduced using CIAO 3.0.1 applying the standard procedure using the latest calibration files.
Source detection was conduced with the wavelet algorithm WAVDETECT included in CIAO. The study was performed in the 0.5-2 keV and 2-10 keV energy bands. Fluxes were estimated assuming as spectral model a power law absorbed by the Galactic column, with the photon index  $\Gamma$ estimated field by field from the stacked spectra of all the sources. The  flux limits were  ~2$\times10^{-16}$ erg cm$^{2}$ s$^{-1}$  and 1$\times10^{-15}$ erg cm$^{2}$ s${-1}$  in the 0.5-2 keV and 2-10 keV bands respectively.
logN-logS were constructed chip by chip with the sky coverage  obtained by imposing that the  fainter detectable source had a detection significance of 3$\sigma$. The logN-logS relations were fitted with a maximum likelihood method assuming as model
a single power-law $k(S/S_{0})^{-\alpha}$. Normalizations ($k$) were computed at flux levels ($S_{0}$) of 2$\times10^{-15}$ erg cm$^{2}$ s$^{-1}$  in the 0.5-2 
keV band and 1$\times10^{-14}$ erg cm$^{2}$ s$^{-1}$  in the 2-10 keV band.
\section{Results}
The logN-logS of sources around the clusters have been calculated chip-by-chip, as well as over the whole 
(4 chips) FOV, and  compared to the logN-logS of the reference fields (the case of 
MS1137+6625 is reported in fig. 1).
 Fig. 2 shows the differential fractional distributions of the logN-logS normalization factor value, $k$, in cluster (shaded area) and reference (white area) fields.
The best-fit normalizations $k$ for the reference fields are centered around the error--weighted means$ k_{soft}$=401$^{+31}_{-13}$ deg$^{-2}$ 
 (standard deviation $\sigma_{soft}$=70$\pm$3 deg$^{-2}$) and  $k_{hard}$=275$^{+22}_{-9}$ deg$^{-2}$  ($\sigma_{hard}$=36$\pm$5 deg$^{-2}$), in
the soft and hard energy bands, respectively.
In six chips of 4 cluster fields,  excesses of a factor $\sim$1.7-2 with a significance $>2\sigma$ are
detected.  The other 6 fields  (24 chips) do not show any significant ($>2\sigma$) excess. 
Nevertheless, in four  out of these six  cluster fields, given the low statistics, factor $\sim2$ 
deviations cannot be statistically  ruled out. 
Figure 2 clearly shows that the distributions have an almost Gaussian shape in the reference fields, both 
in the soft  and hard  energy bands. Instead, in the cluster fields, the distributions 
have a prominent tail at high values of $k$, weighting for about 20-25\% of all cluster fields. 
The significance of the tail  has been quantified with a KS test which yields a probability that the 
two distributions are extracted from the same  parent population of 9.03$\times10^{-3}$ in the soft band 
and 2.55$\times10^{-6}$ in the hard band. 
A very  interesting result obtained from the present analysis is the apparent correlation found 
between the  amplitude of the overdensities  and the cluster redshifts, as shown in Fig. 3. 
In this figure, the ratio $R$ of the (2-10 keV) $k$  normalized to the value of the reference 
fields ($k_{cluster}/k_{reference}$) increases with the cluster redshift. The correlation is significant at a level of $\sim96\%$  
  \begin{figure}
   \centering
     \resizebox{\hsize}{!}{\includegraphics{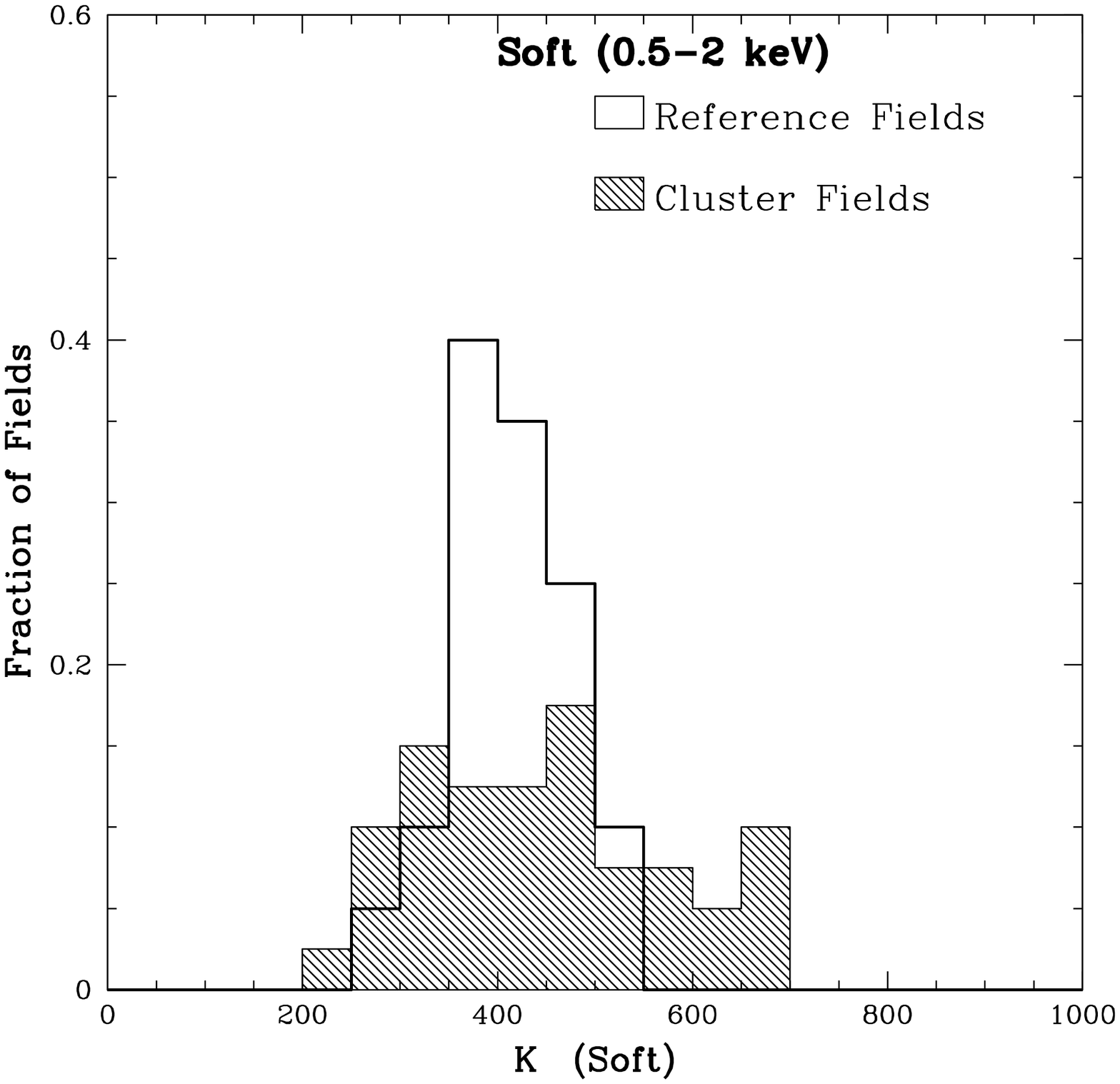}\includegraphics{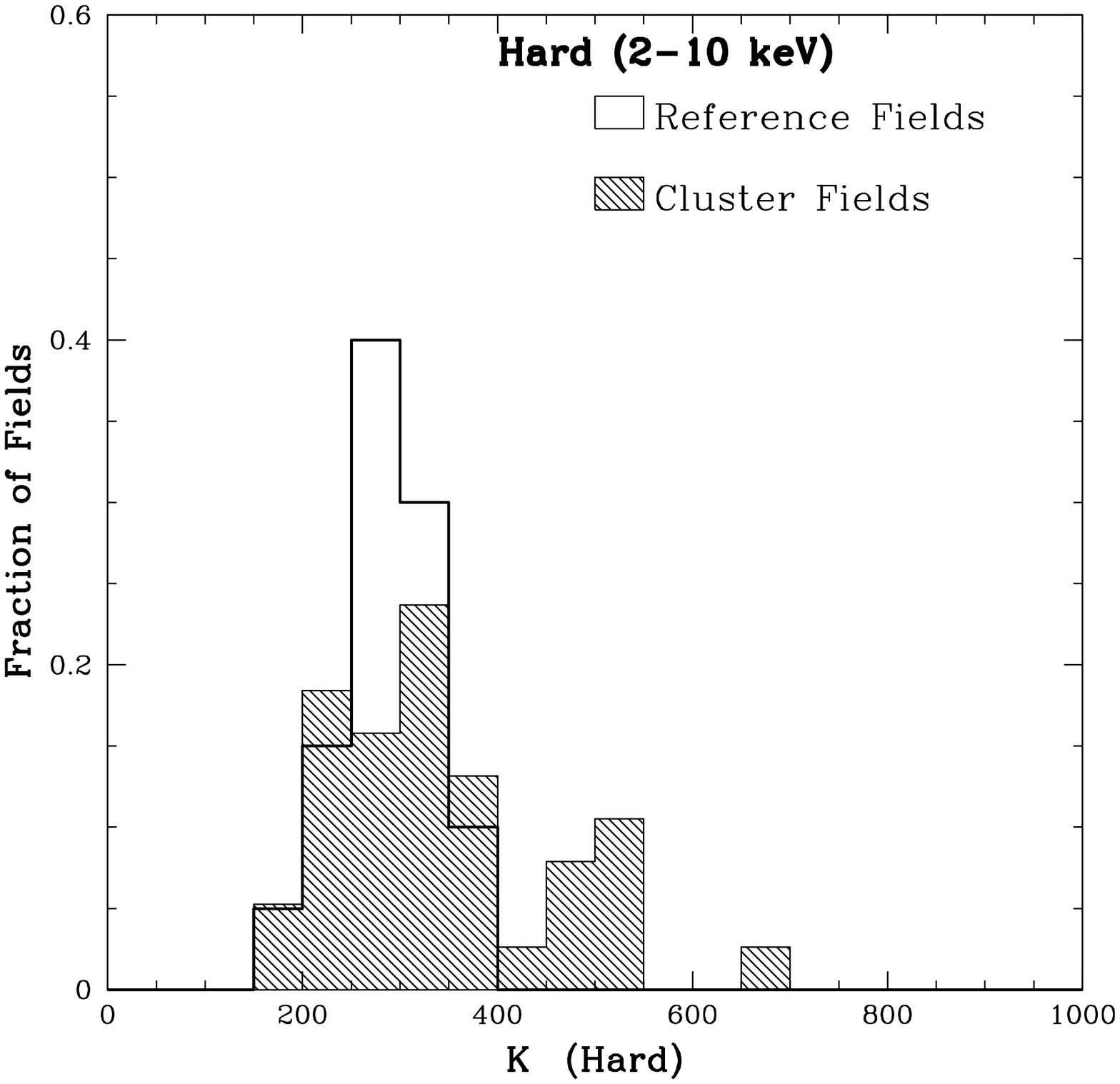}}
      \caption{The frequency histogram of the logN-logS normalization ($K$) in both cluster and non cluster fields (chip-by-chip) in the soft ($Left~Panel$) and hard ($Right~Panel$) energy bands.
              }
         \label{FigVibStab}
   \end{figure}

\section{Conclusions}
 \begin{figure}
   \centering
    \includegraphics[width=6cm,angle=-90]{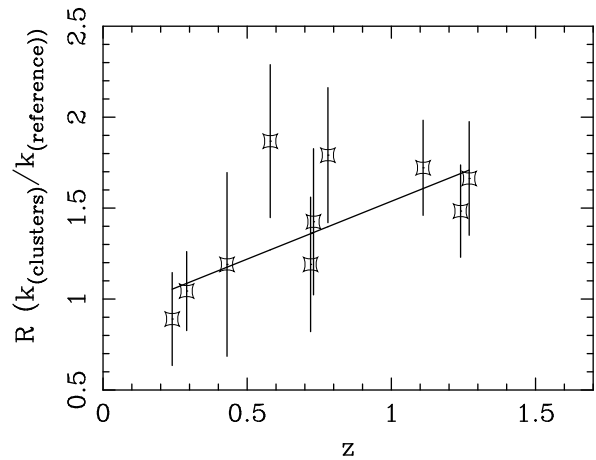}
      \caption{The  ratio R  as function of the cluster $z$ in the hard band.
              }
         \label{FigVibStab}
   \end{figure}
We reported the study of the X-ray source surface density of serendipitous sources around 10 distant 
clusters of galaxies.  4 of these clusters show an X-ray source excess of a factor $\sim$2 with respect to the reference fields. 
This  work brings up to 11 the number of cluster showing this effect.
 If located at the clusters redshift 
these objects should have a luminosity of
10$^{43}$-10$^{44}$ erg/s, typical of low-luminosity QSO and luminous Seyfert galaxies. The rise with the cluster redshift  of the number of sources  is analogous to the Butcher-Oemler effect  in the optical band. 
This would be  the first time it is observed in the X-ray band.  However the excesses here reported are observed on scales 
of 3-7 Mpc (i.e. much larger than the typical dynamical radius of the clusters),  so the Butcher Oemler effect seems not to be an exhaustive explanation here. 
Field--to--field source counts  variations  observed in non cluster fields is typically of the order of 10-20\% (1$\sigma$). 
 A factor of 2 excess cannot therefore be explained with a ``classical'' fluctuation of the  X-ray background.  
Recent  studies conducted on the source excess around the cluster 3C295 clearly show that the 
sources in this field  are strongly clustered (D'Elia et al. 2004). 
We speculate along this line that the most likely explanation of these excesses  can be found following the present knowledge 
of the large scale structure of the Universe:  clusters are observed in the nexus of a web-like  network of filaments. 
Since the sources excesses are observed only on a side of the clusters, and if the AGNs space  
density reflects the galaxies space density, it is possible that we are observing  AGNs in filaments connected with  the clusters. 
In this context the redshift dependence of the excesses could be also explained as a geometrical effect. Under this hypothesis, this study may lead to interesting determination of cosmological  parameters such as $w_{X}$\footnote{ASI and MURST are aknowledged for partial support}.

\bibliographystyle{aa}

\end{document}